\documentclass[sn-mathphys,Numbered]{sn-jnl}% Math and Physical Sciences Reference Style

\usepackage{graphicx}%
\usepackage{multirow}%
\usepackage{amsmath,amssymb,amsfonts}%
\usepackage{amsthm}%
\usepackage{mathrsfs}%
\usepackage[title]{appendix}%
\usepackage{xcolor}%
\usepackage{textcomp}%
\usepackage{manyfoot}%
\usepackage{booktabs}%
\usepackage{algorithm}%
\usepackage{algorithmicx}%
\usepackage{algpseudocode}%
\usepackage{listings}%
\usepackage{cancel}%
\usepackage{bm}
\newcommand{\etal}{{\it et al.,\;}}

\newcommand{\beq}{\begin{equation}}
\newcommand{\eeq}{\end{equation}}
\newcommand{\bea}{\begin{eqnarray}}
\newcommand{\eea}{\end{eqnarray}}

\newcommand{\eF}{\varepsilon_{F}}

\newcommand{\kF}{k_{\textrm{F}}}

\newcommand{\grad}{\boldsymbol{\nabla}} 
\newcommand{\meff}{M_{\textrm{eff}}}
\theoremstyle{thmstyleone}%
\theoremstyle{thmstyletwo}%

\theoremstyle{thmstylethree}%

\usepackage[T1]{fontenc}

\begin{document}

\title[Article Title]{Quantum vortices in fermionic superfluids: from ultracold atoms to neutron stars.}

\author*[1,2]{\fnm{Piotr} \sur{Magierski}}\email{piotr.magierski@pw.edu.pl}

\author[1]{\fnm{Andrea} \sur{Barresi}}\email{andrea.barresi.dokt@pw.edu.pl}
\equalcont{These authors contributed equally to this work.}

\author[1]{\fnm{Andrzej} \sur{Makowski}}\email{andrzej.makowski2.dokt@pw.edu.pl}
\equalcont{These authors contributed equally to this work.}

\author[1,3]{\fnm{Daniel} \sur{P\k{e}cak}}\email{daniel.pecak@pw.edu.pl}
\equalcont{These authors contributed equally to this work.}

\author[1,2]{\fnm{Gabriel} \sur{Wlaz{\l}owski}}\email{gabriel.wlazlowski@pw.edu.pl}
\equalcont{These authors contributed equally to this work.}

\affil*[1]{\orgdiv{Faculty of Physics}, \orgname{Warsaw University of Technology}, \orgaddress{\street{ulica Koszykowa 75}, \city{Warsaw}, \postcode{00-662}, \country{Poland}}}

\affil[2]{\orgdiv{Physics Department}, \orgname{University of Washington}, \orgaddress{\street{3910 15th Ave. NE}, \city{Seattle}, \postcode{WA 98195-1560}, \country{USA}}}

\affil[3]{\orgname{Institute of Physics, Polish Academy of Sciences}, \orgaddress{\street{Aleja Lotnikow 32/46}, \city{Warsaw}, \postcode{02-668}, \country{Poland}}}


\abstract{
Superfluid dilute neutron matter and ultracold gas, close to the unitary regime, exhibit several similarities. Therefore, to a certain extent, fermionic ultracold gases may serve as emulators of dilute neutron matter, which forms the inner crust of neutron stars and is not directly accessed experimentally.
Quantum vortices are one of the most significant properties of neutron superfluid, essential for comprehending neutron stars' dynamics. 
The structure and dynamics of quantum vortices as a function of pairing correlations' strength are being investigated experimentally and theoretically in ultracold gases. Certain aspects of these studies are relevant to neutron stars. We provide an overview of the characteristics of quantum vortices in s-wave-type fermionic and electrically neutral superfluids. The main focus is on the dynamics of fermionic vortices and their intrinsic structure. }


\maketitle

\section{Introduction}\label{sec1}

Pairing correlations in nuclear systems are essential for describing both static and dynamic properties of nuclei. 
Such phenomena as odd-even mass difference or back bending of moments of inertia of atomic nuclei cannot be explained without invoking the concept of pair correlations (see ~\cite{ring2004nuclear, dean2003pairing, brink2005nuclear} and references threrein).
However, the precise extraction of the pairing magnitude is difficult due to the finite size of the system.
The information about pairing gap is obscured by polarization effects associated with
adding or substracting a single nucleon (see eg. \cite{PhysRevC.63.024308}
for discussion of the interplay between pairing and mean-field effects in the odd-even mass staggering).
For most applications, it is sufficient to consider one particular feature of pairing correlations: the presence of the energy gap at the Fermi surface. 
Various manifestations of superfluidity can be traced back to this single quantity, usually treated as a parameter of mean-field models (see e.g. \cite{moller1992nuclear}).
Therefore, it is still an open question whether pairing in nuclear systems can be treated as a condensate of pairs, giving rise to the superfluid phase, or it is merely an incoherent pair correlation.
Clearly, the condition of nonvanishing off-diagonal long-range order cannot be applied to atomic nuclei due to their finite size.
However, if a superfluid phase exists, it has to be seen in phenomena requiring a well-defined phase factor of the pairing field. %\todo{AM: cite Ring\&Schuck?}
These involve analogs of the Josephson effect \cite{goldanskii1968analog, dietrich1970nuclear, dietrich1972semiclassical, magierski2021tiniest}, both in DC  and AC realizations, and solitonic excitations~\cite{PhysRevLett.119.042501, magierski2022pairing}.
Manifestations of these effects were reported to be seen in nuclear collisions
through enhanced neutron pair transfer~\cite{PhysRevC.36.1192, PhysRevC.53.1819, PhysRevC.55.R5}, photon radiation of particular energy~\cite{PhysRevC.93.054623, potel2021quantum, broglia2022nuclear}, or an increase of the barrier for capture
of colliding nuclei~\cite{PhysRevC.97.044611}.

On the other hand, even stronger pairing correlations are predicted to be present in nuclear matter at subnuclear densities~\cite{cao2006screening}. Unfortunately, dilute nuclear, particularly neutron matter, cannot be directly accessed experimentally. They form outer layers of neutron stars --- the so-called inner crust~\cite{chamelHaensel2008}. 
However, since the system is of macroscopic size and is characterized by strong pairing correlations, the superfluid phase (of s-wave type) is likely to be formed~\cite{migdal1959,baym1969superfluidity,pines1985superfluidity,chamel2017}.
Moreover, the dynamics of a rotating star raises the natural question concerning the role of quantum vortices in the dynamics of the crust, especially in the context of observed glitch phenomenon~\cite{anderson1975, haskell2015models}. 
The presence of quantum vortices, which require a particular configuration of pairing field, is possible only under the assumption that the neutron matter is superfluid.

The properties of vortices in neutron stars can be modeled using Density Functional Theory (DFT) based on the properties of known nuclear systems~\cite{PhysRevLett.117.232701, pecak2021properties}. 
However, since neutron matter at subnuclear densities is relatively close to the so-called unitary regime, one can expect that the properties of vortices are close to those observed in ultracold atomic gases.
Therefore, theoretical studies confronted with experiments on ultracold atomic gases can help provide constraints for the theory of quantum vortices in neutron matter~\cite{poli2023glitches}.
Fermionic ultracold gases provide a particularly useful playground to investigate properties of vortices~\cite{giorgini2008theory}, since they provide an example of a system that is electrically neutral, contrary to metal superconductors, which is close to the conditions in the inner crust (proton fraction is small and localized in impurities). 
Therefore, all effects related to interactions with magnetic and electric fields can be ignored.

In this review, we summarize the main characteristics of vortices and their dynamics predicted to exist in neutron matter and those studied in ultracold gases, supported by experiments that are relevant for neutron stars.

\section{Vortex structure and thermodynamic properties}\label{sec2}

The discovery of He-II triggered studies of quantum vortices, which could be rather easily generated in this system. However, in superfluid He-II, the healing length is relatively small, and the vortex core sizes do not exceed a few \AA{ }~\cite{OrtizCeperley}. Consequently, they cannot be directly observed in experiments.
The situation is different in ultracold atomic gases, which are dilute. Therefore, the vortex core size can be about $10^{3}-10^{4}$ times larger, depending on the value of the scattering length.
Attempts to cool down fermionic gases eventually led to the observation of quantum vortices,  which provided unambiguous evidence of superfluidity in the system~\cite{zwierlein2005}.
The regular lattice of vortices has been detected in fermionic gases in a wide range of scattering lengths ranging from Bardeen-Cooper-Schrieffer (BCS) regime through unitary limit to molecular Bose-Einstein Condensate (BEC) regime.
The versatility of ultracold atomic gases allows for considering properties of vortices within the whole range of BEC-BCS crossover. It is particularly important as one can smoothly transform fermionic vortices into bosonic vortices, varying the value of the scattering length.

Both in the BEC and BCS regimes, the presence of a quantum vortex is associated with a particular configuration of the order parameter, which vanishes in the vortex core.
The same is valid for superfluid systems, e.g. He-II, where the "condensate wave function" plays the role of the order parameter.
In both cases, the order parameter is responsible for the stability and dynamics of quantum vortices. 
The main difference, however, between bosonic and fermionic vortices lies in their structure, which, in the latter case, is more complex. In the bosonic systems at $T=0$ the vortex core is essentially empty. 
With increasing temperature, thermal excitations can be generated, forming an admixture of a normal component inside the core.
Regarding fermionic vortices, the particle density in the core does not vanish even at $T=0$.
Indeed, the order parameter in the vicinity of the center of the core creates a particular shape of the pairing potential and does not prevent fermions from occupying the core, leading merely to the emergence of quantized in-gap states.
The first estimates for fermion-bound states in the core have been given in Ref.~\cite{CAROLI1964307}, and therefore, they are sometimes referred to as Caroli-De Gennes-Matricon states.
It was also conjectured that due to these states, thermodynamic properties of superconductors would be affected, namely the specific heat at temperature $k_B T < |\Delta|$ ($|\Delta|$ denotes the magnitude of pairing field far from the vortex) will contain contribution behaving linearly as a function of $T$.
When one goes from a deep BCS regime to the unitary regime, the spectrum of states in the core gets gradually more sparse. The in-gap states are expected to disappear eventually beyond the unitary regime on the BEC side~\cite{PhysRevLett.96.090403}.

Before discussing properties of the core states let us first
summarize briefly Bogoliubov-de Gennes (BdG) formalism, which allow to capture the 
main properties of the quantum vortex.
The explicit form of
BdG equations for spin-imbalanced system reads (no spin-orbit coupling is considered):
\begin{align}\label{eq:hfbspin}
\begin{gathered}
{\cal H} 
\begin{pmatrix}
u_{n,\uparrow}(\bm{r}) \\
u_{n,\downarrow}(\bm{r}) \\
v_{n,\uparrow}(\bm{r}) \\
v_{n,\downarrow}(\bm{r})
\end{pmatrix}
= E_n
\begin{pmatrix}
u_{n,\uparrow}(\bm{r}) \\
u_{n,\downarrow}(\bm{r}) \\
v_{n,\uparrow}(\bm{r}) \\
v_{n,\downarrow}(\bm{r})
\end{pmatrix}, \\
{\cal H} =  
\begin{pmatrix}
h_{\uparrow}(\bm{r}) -\mu_{\uparrow}  & 0 & 0 & \Delta(\bm{r}) \\
0 & h_{\downarrow}(\bm{r}) - \mu_{\downarrow}& -\Delta(\bm{r}) & 0 \\
0 & -\Delta^*(\bm{r}) &  -h^*_{\uparrow}(\bm{r}) +\mu_{\uparrow} & 0 \\
\Delta^*(\bm{r}) & 0 & 0& -h^*_{\downarrow}(\bm{r}) + \mu_{\downarrow}
\end{pmatrix},
\end{gathered}
\end{align}
where $\mu_{\uparrow,\downarrow}$ are chemical potentials for spin-up and spin-down
particles, respectively.
Single particle hamiltonian $h_{\uparrow}=h_{\downarrow}=-\frac{\hbar^2}{2m}\nabla^2$
consists of the kinetic term and the mean-field potential, but
we will omit the latter, as of the secondary importance in this case.
In principle, the single particle hamiltonian $h_{\sigma}$
may explicitly depend on the spin state (see eg. Ref.~\cite{bulgac2012}).
The pairing gap is related to quasi-particle wave-functions:
\begin{equation}
\Delta(\bm{r}) = -\dfrac{g_{\textrm{eff}}}{2}\sum_{0<E_n<E_c} (u_{n,\uparrow}(\bm{r})v_{n,\downarrow}^{*}(\bm{r})-u_{n,\downarrow}(\bm{r})v_{n,\uparrow}^{*}(\bm{r})),\\
\end{equation}
where $g_{\textrm{eff}}$ is a regularized coupling constant and $E_c$ is cut-off energy scale, see~\cite{bulgac2012} for details of the regularization scheme.

The BdG equations~(\ref{eq:hfbspin}) decouple into two independent sets:
\begin{align}\label{eq:hfbspin1}
\begin{gathered}
\begin{pmatrix}
h_{\uparrow}(\bm{r})-\mu  &  \Delta(\bm{r}) \\
\Delta^*(\bm{r}) &  -h^*_{\downarrow}(\bm{r})+\mu
\end{pmatrix} 
\begin{pmatrix}
u_{n,\uparrow}(\bm{r}) \\
v_{n,\downarrow}(\bm{r})
\end{pmatrix}
= E_{n+}
\begin{pmatrix}
u_{n,\uparrow}(\bm{r}) \\
v_{n,\downarrow}(\bm{r})
\end{pmatrix} ,
\end{gathered}
\end{align}
\begin{align}\label{eq:hfbspin2}
\begin{gathered}
\begin{pmatrix}
h_{\downarrow}(\bm{r})-\mu& -\Delta(\bm{r}) \\
-\Delta^*(\bm{r}) &  -h^*_{\uparrow}(\bm{r})+\mu
\end{pmatrix} 
\begin{pmatrix}
u_{n,\downarrow}(\bm{r}) \\
v_{n,\uparrow}(\bm{r})
\end{pmatrix}
= E_{n-}
\begin{pmatrix}
u_{n,\downarrow}(\bm{r}) \\
v_{n,\uparrow}(\bm{r})
\end{pmatrix},
\end{gathered}
\end{align}
where $\mu=\frac{1}{2}(\mu_{\uparrow}+\mu_{\downarrow})$ denotes mean chemical potential and $E_{n\pm}=E_n \pm \frac{\Delta\mu}{2}$ with $\Delta\mu=\mu_{\uparrow}-\mu_{\downarrow}$.
Solutions of equations (\ref{eq:hfbspin1}) and (\ref{eq:hfbspin2}) are connected through 
the symmetry relation, namely if vector $\varphi_{+}=\left( u_{n\uparrow},v_{n\downarrow}\right)^{T}$ represents a solution of Eq.~(\ref{eq:hfbspin1}) with eigenvalue $E_n$, then vector $\varphi_{-}=(v_{n\uparrow}^*,u_{n\downarrow}^*)^{T}$ is a solution of Eq.~(\ref{eq:hfbspin2}) with eigenvalue $-E_n$.
Therefore it is sufficient to solve  equations (\ref{eq:hfbspin1}) only (for all quasiparticle energy states), and then solutions with positive quasiparticle energies contribute to the spin-down densities, whereas solutions with negative energies to the spin-up densities.

For the solution describing a vortex the particular form of pairing field is required
$\Delta(\pmb{r}) = |\Delta(r)|\exp(i\phi)$ (for the vortex rotating clockwise), 
where $(r, \phi, z)$ are cylindical coordinates
with $r$ being the distance from the center of the vortex core and $z$ - the coordinate
along the vortex line (which is assumed to be a straight line). Therefore the pairing field does
not depend on $z$ and the general solution of eq.~(\ref{eq:hfbspin}) describing the vortex reads:
\begin{align} \label{coresolution}
\begin{gathered}
\begin{pmatrix}
u_{n,\uparrow}(\bm{r}) \\
v_{n,\downarrow}(\bm{r})
\end{pmatrix}
= 
\begin{pmatrix}
u_{n m k_{z},\uparrow}(r) e^{i m \phi} e^{i k_{z} z} \\
v_{n m k_{z},\downarrow}(r) e^{i (m-1) \phi} e^{i k_{z} z}
\end{pmatrix} ,
\end{gathered}
\end{align}
The other pair of solutions can be obtained using the symmetry properties of
BdG equations. Note that (\ref{coresolution}) implies
that the angular momentum of spin-up particles (holes) differs from
spin-down holes (particles), of the same energy, by $\hbar$.
%%%%%%%%%%%%%%%%%%%%%%%%%%%%%%%%%%%%%%%%%%%%%%%%

The spectrum of quantized states inside the vortex core has a peculiar structure, reflecting the behavior of the order parameter. 
In the deep BCS limit, the main features of the spectrum can be reproduced within the semiclassical Andreev approximation~\cite{andreev1964}.
In this approximation one assumes the separation of scales related to the density and the pairing field spatial modulation. The former is set by the Fermi wavelength, whereas
the latter - by the coherence length $\xi$.
Therefore one may decompose the variation of $u$ and $v$ components of
wave-functions (see Eq.~(\ref{eq:hfbspin})) at the Fermi surface into rapidly oscillating parts associated with the Fermi momentum $\kF$ and smooth variations governed by the coherence length, i.e.~$u(\bm{r})= e^{i\bm{k}_\textrm{F}\cdot\bm{r}}\tilde{u}(\bm{r})$ with $|\bm{k}_\textrm{F}|=\kF$, and similarly for the $v$ component.
The Andreev approximation can be used also in the case of spin imbalance systems, providing the local polarization is relatively weak $\Delta\mu = \mu_{\uparrow}-\mu_{\downarrow} \ll \frac{1}{2}(\mu_{\uparrow}+\mu_{\downarrow}) \approx \eF= \kF^{2}/2 $.
We will focus only on one set of BdG equations, which in Andreev
approximation describing states close to the Fermi surface acquire the form (we set $\hbar=m=1$):
\begin{equation} \label{bdg1}
 \left( \begin{array}{cc}
 -i\bm{k}_\textrm{F}\cdot\grad& \Delta(\textbf{r}) \\
 \Delta^*(\textbf{r}) & i\bm{k}_\textrm{F}\cdot\grad
\end{array} \right)\left( \begin{array}{cc}
\tilde{u}_{n,\uparrow}(\textbf{r}) \\
\tilde{v}_{n,\downarrow}(\textbf{r})
\end{array} \right) =\tilde{E}_{n+} \left( \begin{array}{cc}
\tilde{u}_{n,\uparrow}(\textbf{r}) \\
\tilde{v}_{n,\downarrow}(\textbf{r})
\end{array} \right),
\end{equation}
where $\tilde{E}_{n+} = E_{n+} + \frac{\Delta\mu}{2}$.
The second pair of equations for $\tilde{u}_{n,\downarrow}(\textbf{r})$ and $\tilde{v}_{n,\uparrow}(\textbf{r})$ has similar form and correspond to $\tilde{E}_{n-} = E_{n-} - \frac{\Delta\mu}{2}$.

\begin{figure}[h]
   \begin{center}
   \includegraphics[width=1.0\textwidth]{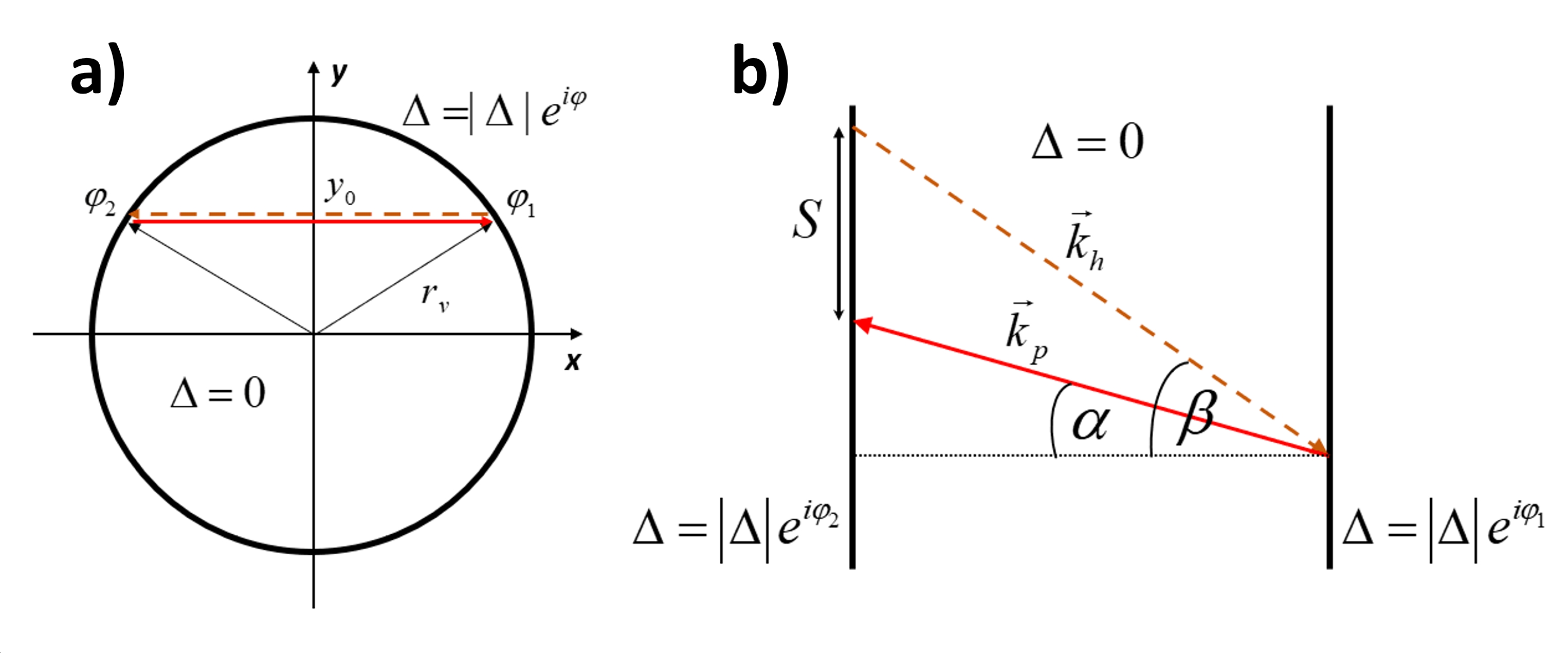}
   \centering
   \end{center}\vspace{-3mm}
   \caption{A schematic picture of sections through the vortex core.
   Section perpendicular to the vortex line is shown in the left subfigure a).
   The classical trajectory representing particle of momentum $k_{F}$ is denoted 
   by a red solid line, and the retroreflected hole is shown as a brown dashed line. 
   Note that the angular momentum component $L_{z}$ corresponding to the trajectory is 
   negative, and the vortex rotates counterclockwise.
   Section along to the vortex line is shown in the right subfigure b).
   The classical trajectory representing particle 
   of momentum $k_{p}$ is denoted by a red solid line, and the reflected hole of 
   momentum $k_{h}$ is shown as a brown dashed line.
   } \label{fig1}
\end{figure}

%%%%%%%%%%%%%%%%%%%%%%%%%%%%%%%%%%%%%%%%%%%
Although, Andreev approximation requires that $|\Delta|/\eF \ll 1$ ($\eF$ denotes the Fermi energy), the spectrum of low-lying states can be qualitatively reproduced using a schematic model with pairing field defined as $\Delta(r,\phi)=|\Delta|e^{i\phi}\theta(r-r_{v})$
and shown in Fig. \ref{fig1}(a). The reason is that, the most important ingredient of
the model is related to the particular phase pattern of the pairing potential, whereas
the asymptotic behavior of the wave functions play only a minor role (for a comprehensive
discussion of the quality of Andreev approximation see Ref.~\cite{adagideli2002}).

\begin{figure}[h]
 \begin{center}
 \includegraphics[trim={0 0 0 32mm},clip,width=0.75\textwidth]{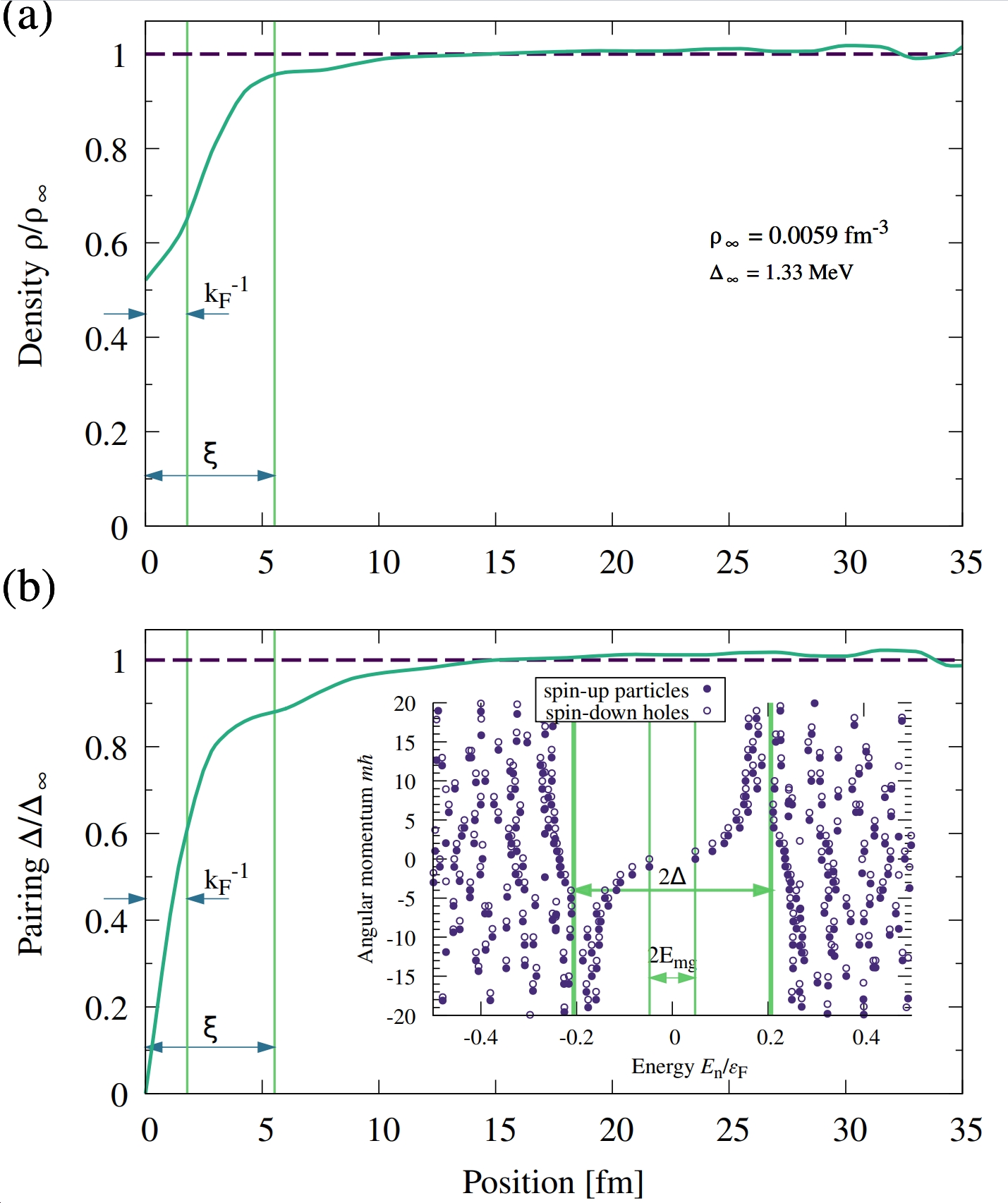} 
 \end{center}
  \caption{The cross-section through the center of the vortex in neutron matter of bulk density $\rho_{\infty} = 0.0059$ fm${{}^{-3}}$ and corresponding bulk pairing gap $\Delta_{\infty} = 1.33$ MeV. The parameters correspond to the inner crust region of a neutron star. The cross-section of (a) density and pairing field (b) (both quantities normalized to the corresponding bulk value). Two characteristic length scales are shown: inverse Fermi momentum $k_F^{-1}$ and the superfluid coherence length $\xi$. Inset in (b): the energy spectrum of quantum states (of particle and hole types) and their quantized angular momenta (the relative shift of angular momenta 
 between particles and holes
is the consequence of relation (\ref{coresolution})). The energy scales governing the physics of the system --- bulk energy gap $\Delta_{\infty}$, as well as so-called minigap $E_{mg}$ --- are shown. The results are obtained with DFT framework for nuclear systems~\cite{pecak2021properties}.
  }
  \label{fig2}
\end{figure}

Then, the quantization condition, originating from the perfect particle-hole
retroreflection, reads~\cite{PhysRevA.106.033322}:
\begin{eqnarray}
\label{quant}
\frac{E_{n}}{\eF}\kF r_{v}\sqrt{1-\left (\frac{L_{z}}{\kF r_{v}} \right )^{2}}+\arccos{\left ( \frac{-L_{z}}{\kF r_{v}} \right )}
- \arccos{\frac{E_{n}}{|\Delta|}}=\pi n,
\end{eqnarray}
where $n\in\{0,\pm 1, \pm 2,\dots$\}, $r_{v}$ denotes the radius of the vortex core, $\kF$ - Fermi momentum, and $|L_{z}|=r\kF$.
Note that only the states with $n=0$ correspond to core states, i.e.~$E_{n=0}\lesssim |\Delta|$\footnote{While solving the full BdG eqs. the states corresponding to $n \neq 0$ can still be found at energies smaller than $|\Delta|$, but they are very close to continuum.}.
The limit $|E|\ll |\Delta|$ can be quite accurately approximated by the expression:
\begin{equation} 
\label{andreev_pol}
E_{\pm, n=0,m} \approx -\frac{|\Delta|^{2}}{\eF\frac{r_{v}}{\xi}\left (\frac{r_{v}}{\xi} + 1 \right )}  m ,
\end{equation} 
where $m$ is the magnetic quantum number associated with $L_{z}=\hbar m$, pointing along the vortex axis and $\xi=\eF/(\kF |\Delta|)$ is a coherence length. Minus sign in front of the rhs expression (\ref{andreev_pol}) is related to the counterclockwise rotation of the vortex.
Linear dependence $E\propto L_{z}$ holds until the energy approaches $\Delta$, and subsequently, it is bent towards large angular momenta. This part of the spectrum cannot be reproduced within Andreev approximation.
In the regime close to unitarity and the inner crust of neutron stars, the linear part of the spectrum consists of a few states only, as shown in the inset of Fig.~\ref{fig2}.
The results presented in this figure have been obtained by solving numerically
Hartree-Fock-Bogoliubov equations (\ref{eq:tdslda2c}) 
(having the same form as eqs. (\ref{eq:hfbspin})) on the spatial lattice (in 3D) for the pure neutron matter. The BSk energy density functional with the local pairing field $\Delta(\pmb{r})$ has been
used (see Section 3). 
The resulting profile of the pairing field and 
the neutron density distribution $\rho(\pmb{r})$,
have been obtained selfconsistently through the total energy minimization, 
with the requirement that the pairing
field possess the structure characteristic for the vortex located in the center of 
the cylindrical box. Thus the results represent more realistic structure of the vortex core.
The quasiparticle energies as a function of the angular momentum along the vortex symmetry
axis have been plotted in the inset. The plotted energies correspond to $k_{z}=0$
in expression (\ref{coresolution}).
%%%%%%%%%%%%%%%%%%%%%%%%%%%%%%%%%%%%%%%%%%%%%%%%%%%

%%%%%%%%%%%added/corrected%%%%%%%%%%%%%%%%
Note that in the inner crust of the neutron star, in the region where the s-wave pairing is
predicted to be the strongest, i.e. with the value of the pairing gap 
within the interval $(1.2, 1.6)$ MeV,
the size of the vortex core is much smaller (at $T\lesssim 0.1$ MeV) 
than the typical modulation of the nuclear density, induced by the presence of 
nuclear lattice. Namely, the radii of the
vortex cores do not exceed $12$ fm~\cite{pecak2021properties}, whereas typical
lattice constants in this region range between $40$ and $70$ fm \cite{refId0}.
Therefore, for the nuclear densities $(0.006 - 0.034)$ fm$^{-3}$ it 
is possible to place the vortex core in the region, where the matter is uniform.
The situation is different, however, if one approaches the boundaries of the inner
crust. In the limit of small densities, at the border of the outer
crust region, the lattice constant reaches values of about $100$ fm, whereas 
the size of the vortex cores can be arbitrarily large due to the vanishing density 
of the neutron gas outside nuclei. The similar situation occurs
in the region, where pasta phases are predicted to exist, i.e. at the densities $0.05$ fm$^{-3}$
and higher. The typical length scale of nuclear density modulation is about $20$ fm 
and the vortex core can be of similar size. In these cases one has to take into account that
the density inside the vortex core is in addition modulated due to the particular nuclear structure, independent of the shape of the pairing field associated with the vortex.
This effect will in turn introduce additional modification of the structure of the
in-gap states, due to the scattering of quasiparticles on nuclear inhomogeneities.
Moreover, in these region, the scattering of quasiparticles (inside the vortex core) on nuclear inhomogeneities may provide another efficient source of energy dissipation 
for a moving vortex (see discussion in Section 3).
%%%%%%%%%%%%%%%%%%%%%%%%%%%%%%%%%%%%%%%%%%

There are few immediate consequences to the existence of the so-called chiral branch corresponding to $n=0$.
Since the quasiparticle excitation spectrum consists of states with $m < 0$ (for a vortex rotating counterclockwise), the energy of the first excited state $E_{mg}\approx \frac{|\Delta|^2}{2\eF}$ (assuming $r_v=\xi$) sets the energy scale (minigap energy) associated with the excitation of the vortex core.
Moreover, the chiral branch's structure determines the density behavior in the vortex core. Specifically, since the $m=0$ state is unoccupied, the density reveals depletion in the core. 
It can be deduced from BdG eqs.
that density $\rho(r) \propto r^2 + \textrm{const}$ in the vicinity of the center of the vortex.
The vortex core structure, obtained within the framework 
of Density Functional Theory (see section \ref{sec3}), is shown in Fig.~\ref{fig2}. It has to be emphasized that the depletion of the density inside a vortex core is crucial for the experimental detection of vortices in ultracold Fermi gases\footnote{It has to be mentioned, that the depletion in density profiles,
in the presence of the vortex, have been confirmed in ab-initio Quantum Monte Carlo approach
\cite{PhysRevA.95.053603,PhysRevC.100.014001}, although the accuracy of these results is
affected by the finite size effects.}.

The presence of a chiral branch has an impact on thermodynamic properties as well. For temperatures within the range $E_{mg} \ll k_B T \ll |\Delta|$, it was predicted~\cite{CAROLI1964307} that the specific heat should be a linear function of temperature. It is, however, true in deep BCS limit only ($\Delta\ll\eF$). Due to the limited number of in-gap states in strongly interacting systems, the temperature range fulfilling the above criterion
may not exist. Moreover, the dependence of the size of the core on temperature makes the functional dependence of the specific heat more complicated~\cite{pecak2021properties}.
Another modification of the specific heat comes from the superflow surrounding a vortex due to the rearrangement of the quasiparticle spectrum induced by the flow.
Namely, the quasiparticle dispersion relation in the presence of superflow with velocity $v_{s}$ reads:
\begin{equation}
E_{\pm}(\pmb{k},\pmb{v_s}) = 
\hbar\pmb{k}\cdot\pmb{v_s}\pm \sqrt{\left ( \frac{ \hbar^{2}k^{2}}{2M} - 
\tilde{\mu} \right)^{2} + |\Delta|^{2} },
\end{equation}
where $\tilde{\mu} = \mu -\frac{1}{2}M v_{s}^{2}$ represents the correction
to the chemical potential due to the superflow.
We consider the case $\hbar \kF v_{s}/|\Delta| \ll 1$ corresponding to velocities
smaller than the critical velocity. In that case $E_{+}(\pmb{k},\pmb{v_s}) > 0$
and one can determine the correction to the specific heat for the moving superfluid.
Keeping terms up to $v_{s}^{2}$ one gets:
\begin{align}
&C_{V}(v_{s}) \approx\frac{2}{T^2}V N(0) \sqrt{2\pi T|\Delta|} \exp\left ( -\frac{|\Delta|}{T} \right ) \nonumber \\
&\times \left [ |\Delta|^{2}+\frac{\mu Mv_{s}^{2}}{3} 
\left ( \left (\frac{|\Delta|}{T}\right )^{2}-4 \frac{|\Delta|}{T}+2-\frac{3}{4}\left (\frac{|\Delta|}{\mu}\right )^{2} \right )^{2} \right ],
\end{align}
where 
\begin{equation}
N(0)=\frac{1}{(2\pi)^{2}}\left ( \frac{2M}{\hbar^{2}} \right )^{3/2}\sqrt{\mu}
\end{equation}
is the density of states at the Fermi surface. 
The last term, proportional to $\left (|\Delta|/\mu \right )^{2}$, is a correction due to the modification of the chemical potential $\tilde{\mu}$ in the presence of superflow. It is, however, at least an order of magnitude smaller than the other 
terms and can be neglected. One may
apply this formula to the flow induced by a vortex within the local density
approximation. Namely, one can divide 
the volume around the vortex line into infinitesimal concentric cylindrical shells (of length $L$), in which the superfluid velocity is constant,
and subsequently integrate the contributions to the specific heat coming
from each shell. For the cylinder of radius $R_{out}$
one gets the correction to specific heat per unit length of the vortex: 
\begin{align}
\frac{\Delta C_{V}^{flow}}{L} \approx &\frac{1}{3} (2\pi)^{3/2} N(0)\sqrt{\frac{|\Delta|}{T}}\frac{\mu}{T}
\frac{\hbar^2}{2M} \ln\left ( \frac{R_{out}}{r_{v}}\right )  
\left [ \left (\frac{|\Delta|}{T}\right )^{2}
-4 \frac{|\Delta|}{T} + 2 \right ] \exp\left ( -\frac{|\Delta|}{T} \right ) .
\end{align}
As discussed in Ref.~\cite{pecak2021properties}, $R_{out}$ is of the order of inter-vortex distance in the case of vortex lattice.

%%%%%%%%%%%%%added/corrected%%%%%%%%%%%%%%%
Consequently, the modification of the specific heat per unit volume, due to the presence
of vortices, reads:
\begin{equation}
\frac{1}{V}\Delta C_{V} = \sigma_{vor}\left ( \frac{\Delta C_{V}^{core}}{L} +
\frac{\Delta C_{V}^{flow}}{L}  \right ),
\end{equation}
where $\sigma_{vor} = N_{vor}/S$ (and $V=S\times L$) is the surface vortex density and the two terms:  $\Delta C_{V}^{core}$,
$\Delta C_{V}^{flow}$ represent modification of the specific heat due to the vortex core and
the flow induced by the vortex, respectively.
Due to small average vortex density $\sigma_{vor}\approx 10^{-21}$ fm$^{-2}$, 
the modification of the specific heat of the neutron star crust is negligible for the temperatures 
$T \gtrapprox 0.1$ MeV. However, if the vortex lattice is perturbed, giving rise to the turbulent-like
behavior, one can expect that the vortex density will exhibit significant spatial
fluctuations~\cite{wlazlowski2024}. Using the results of Ref.~\cite{pecak2021properties}
(see Fig. 10 therein) one can estimate that the local increase of vortex
density to $\sigma_{vor}\approx 10^{-2}$ fm$^{-2}$ leads to approximately twice larger
specific heat at $T\approx 0.1$ MeV for a wide range of neutron densities in the crust 
$(0.006 - 0.034)$ fm$^{-3}$. For even lower $T$, specific heat will increase exponentially.
%%%%%%%%%%%%%%%%%%%%%%%%%%%%%%%%%%%%%%%%%%%%%%

The core states, discussed above, correspond to 2D vortex, i.e., they are characterized by $k_{z}=0$,
where $k_{z}$ is a momentum along the vortex line. 
In 3D, corresponding to the infinite vortex line, each state forming the chiral branch represents a band. The slope of the band is directly related to the effective mass of quasiparticle excitations, and it can be traced back to the quasiparticle scattering properties along the vortex line. 
In Fig. \ref{fig1}(b), the schematic picture of Andreev scattering,
leading to the motion of quasiparticles along the vortex line, has been shown.
Contrary to the quantization condition, which resulted from the assumption that the hole(particle) is reflected exactly backward (which is true if the incoming particle(hole) is at the Fermi surface), here we need to release this constraint to determine the effective mass.
Consequently one needs to take into account
corrections related to $\eF \pm E$.
As a result of the momentum conservation along the vortex line, the reflection law reads: $\sqrt{\eF + E}\sin\alpha = \sqrt{\eF - E}\sin\beta$, where $k_{p}=\sqrt{2(\eF + E)}$ and $k_{h}=\sqrt{2(\eF - E)}$ are particle and hole momenta, respectively. 
The effective velocity along the vortex line can be evaluated as $v_{z}=\frac{S}{\Delta T}=\sqrt{2(\eF + E)}\sin\alpha\sin(\beta-\alpha)/\sin(\beta+\alpha)$, where $\Delta T$
denotes the time interval between two consecutive particle-to-hole reflections.
Finally one gets~\cite{PhysRevA.106.033322}:
\begin{equation}
v_{z}=k_{z}\frac{\sqrt{k_{p}^{2}-k_{z}^2}-\sqrt{k_{h}^{2}-k_{z}^2}}{\sqrt{k_{p}^{2}-k_{z}^2}+\sqrt{k_{h}^{2}-k_{z}^2}},
\end{equation}
where $k_{z}=k_{p}\sin\alpha=k_{h}\sin\beta$ is the momentum component along the vortex line. 
Considering the linear term in $k_{z}$ and $E$ on the rhs, one obtains the effective mass $\meff^{-1}\approx E/2\eF$, which agrees with the effective mass derived from 
the formula for the dispersion relation in the BCS limit $E(k_{z})=E(0)/\sqrt{1-k_{z}^{2}/(2 \eF)}$~\cite{CAROLI1964307}.
One may also estimate the magnitude of the effective mass component 
along the vortex line, corresponding to angular momentum $L_{z}=\hbar m$: $\meff^{-1}(m)\approx\frac{2|m|}{3}\left (\frac{\Delta}{\eF} \right )^{2}$.
Note that, in deep BCS limit, the inverse of the effective mass will be exponentially 
small since $\Delta/\eF\propto e^{-\pi/2|a|\kF}$, and clearly, the departure from the flat band behavior will be significant at the unitarity, where $\Delta/\eF\approx 0.5$.
The band flatness, increasing the effective mass, will affect the propagation of the confined quasiparticle excitations, e.g., in the form of local spin-polarization, along the vortex line.
It was shown in Ref. \cite{tylutki2021universal} that the locally induced spin-polarization in the vortex core (e.g., due to the reconnection process with spin-polarized vortex) can hardly propagate along the vortex line.
Indeed, the motion along the vortex line is characterized by the velocity $v_{z}=k_{0}/\meff\propto k_{0}\left (\frac{\Delta}{\eF} \right )^{2}$, where $k_{0}$ is the initial momentum of the wave packet.
Similarly, the wave packet width, in the limit of long times, behaves as $\sqrt{\langle (z - v_{z}t)^{2} \rangle} \propto t\left (\frac{\Delta}{\eF} \right )^{2}$ and leads to an effective suppression of the polarization propagation, see Ref. \cite{PhysRevA.106.033322}.

\begin{figure}[h]
   \begin{center}
   \includegraphics[width=0.55\textwidth]{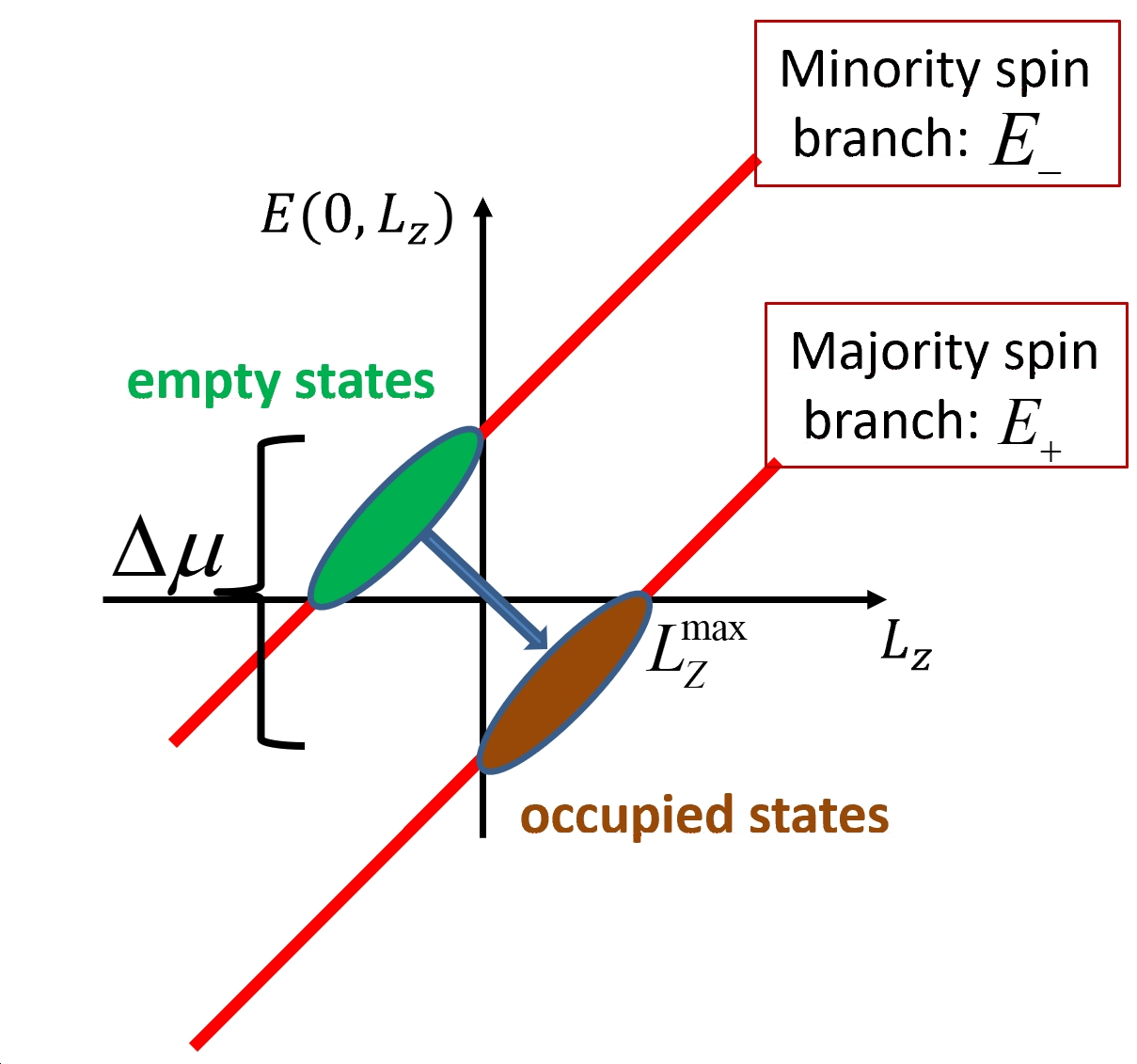}
   \centering
   \end{center}\vspace{-3mm}
   \caption{Chiral branches (red solid line) of low-lying in-gap states in the vortex core for clockwise rotating vortex. Parts of branches that became empty and occupied due to induced spin imbalance have been shown in green and brown colors, respectively.}
   \label{fig4}
\end{figure}

The structure of the vortex core is sensitive to spin imbalance. 
When the chemical potentials for spin-up ($\mu_{\uparrow}$) and spin-down ($\mu_{\downarrow}$) fermions begin to differ, the core of the vortex will be affected first, before the superfluid will be modified in bulk. This is %clearly 
due to the fact that $E_{mg} < |\Delta|$.
Indeed, unpaired fermions tend to accumulate at the core, as shown in~\cite{takahasi2006,drummond2007}.
Consequently, the chiral branch splits into two components corresponding to spin-up and spin-down particles.
The main features of this splitting can be described 
introducing a slight modification of 
the expression (\ref{quant}), which reads:
\begin{eqnarray}
\frac{\tilde{E}_{n\pm}}{\eF}\kF r_{v}\sqrt{1-\left (\frac{L_{z}}{\kF r_{v}} \right )^{2}}+\arccos{\left ( \frac{-L_{z}}{\kF r_{v}} \right )}  
- \arccos{\frac{\tilde{E}_{n\pm}}{|\Delta|}}=\pi n,
\end{eqnarray}
where $\tilde{E}_{n\pm} = E_{n\pm} \pm \frac{\Delta\mu}{2}$ and $\Delta\mu=\mu_{\uparrow}-\mu_{\downarrow}$ is the difference between chemical potentials of spin-up and spin-down particles, respectively. A schematic representation of the solutions corresponding to the chiral branch ($n=0$) is shown in Fig.~\ref{fig4}.
The immediate consequence of this splitting results in occupying states, by majority spin particles, which rotate in opposite directions than the vortex. The highest value of angular momentum corresponding to the opposite rotation $L_{z}^{max}=\hbar m_{opposite}$ is related to spin imbalance and reads:
\begin{equation}
\max|m_{\textrm{opposite}}|\approx \frac{1}{2}\frac{\eF}{|\Delta|^2}\frac{r_{v}}{\xi}\left (\frac{r_{v}}{\xi} + 1 \right )\Delta\mu.
\end{equation}
It is, therefore, clear that the reversed flow of %the majority of spin 
these particles partly cancels the flow of the unpolarized component. 
However, as was shown in Ref.\cite{PhysRevA.106.033322}, the cancellation is not exact and leads to a reversal of the total flow in the core.
The splitting of the chiral branch, induced by the spin imbalance, 
which rearranges the occupation of low-lying states, has another effect. Namely, the $m=0$ state becomes now occupied, and the density is no longer depleted inside a polarized vortex. It makes the experimental detection of vortices in spin imbalanced system far more complex.

In the case of quantum vortices in the inner crust of a neutron star, the mechanism that may lead to the polarization of the vortex core is due to the magnetic field.
It couples to the neutron magnetic moment and may induce a spin flip. 
In Ref.~\cite{pecak2021properties} the minimal value of the magnetic field required to produce such 
an excitation in the vortex core has been estimated to range between about $10^{15}-10^{16}G$, depending on neutron density. 
Note, that this value is more than an order of magnitude smaller than magnetic field
needed to destroy superfluidity in the neutron star crust,
which was estimated as $B\ge 10^{17}G$~\cite{PhysRevC.93.015802}.
There are known magnetars where such conditions may exist~\cite{kaspi2017magnetars}.
%%%%%%%%%

The important effect of polarization in the vortex core is that it eventually causes the minigap to vanish. As was shown in Ref.~\cite{PhysRevA.106.033322}, this is due to the fact that the spectrum of states with $k_{z}=0$ is asymmetric with respect to Fermi surface (see Fig. \ref{fig4}). However, the spectrum of states corresponding to $k_{z}\gg \kF$ becomes symmetric, which is apparent by inspecting the structure of the BdG Hamiltonian, describing the infinite vortex line.
Therefore, one may infer that at certain values of $k_{z}=\pm k_{z1}, \pm k_{z2},\dots$ the quasiparticle energies vanishes $E(\pm k_{zi})=0$.
When the quasiparticle energy changes from negative to positive value, the particle state $v_{\uparrow}$ with momentum $m$ is converted into the hole state $u_{\uparrow}$ with momentum $-m+1$, inducing
the change of $m$: $\Delta m = |2m-1|$.
Therefore, the spin-imbalanced vortex is characterized by a series of quasiparticle level crossings at the Fermi surface, i.e., points at which the minigap vanishes.
Eventually, if the spin imbalance increases, it is predicted that modulation of the pairing field, similar to the Fulde-Ferrell-Larkin-Ovchinnikov (FFLO) phase, will be seen~\cite{PhysRevA.103.053308}.

All these characteristics of the fermionic vortex core structure should have an impact on dynamics and, in particular, on dissipative processes associated with the vortex motion. These will
be discussed in the next section.

\section{Vortex dynamics}\label{sec3}

The problem of vortex dynamics is essential for understanding the behavior of superfluids, particularly the role of dissipative processes in the decay of quantum turbulence~\cite{tsubota2008quantum, haskell2020turbulent}.
The apparent differences between the structure of bosonic and fermionic vortices, discussed in the previous section, raise a question about the impact on vortex dynamic.
The existence of another energy scale associated with minigap and the presence of the chiral branch is expected to generate additional contributions to the dissipative force. 
Last but not least, there is still an open question associated with vortex inertia, which in the case of bosonic vortices is usually assumed to be negligible~\cite{nakamura2012dynamics}.

Answers to these questions should result in formulating the effective equation of motion of a fermionic vortex, where all terms will acquire microscopic underpinning.
If we consider a 2D vortex, i.e., neglecting degrees of freedom associated with the susceptibility of vortex lines to bending and generating Kelvin waves, the problem is still complex as 
the most general equation of motion comprises the following terms:
\begin{equation} \label{eqmotion}
M_{V}\frac{d^{2}\pmb{R}_{V}}{dt^{2}}=\gamma\frac{d\pmb{R}_{V}}{dt} + \omega\frac{d\pmb{R}_{V}}{dt}\times\pmb{e}_{z} + \pmb{F}_{\textrm{superfluid}} + \pmb{F}_{\textrm{pinning}},
\end{equation}
where  $\pmb{R}_{V}$ is the vortex position,  $\pmb{e}_{z}$ is the unit vector perpendicular to the plane of vortex motion, $\pmb{F}_{\textrm{pinning}}$ describes the interaction with inhomogeneities and $\pmb{F}_{\textrm{superfluid}}$ is the component of the force acting on a vortex but independent of its velocity.
The equation arises from the general expression of forces that can be present in the system 
and being in agreement with Galilean invariance. The latter requirement implies that
the force either depends on $(\frac{d\pmb{R}_{V}}{dt} - \pmb{v}_{s} )$ or on $(\frac{d\pmb{R}_{V}}{dt} - \pmb{v}_{n} )$, where $\pmb{v}_{s}$ and $\pmb{v}_{n}$ describe the velocity of superfluid and normal components, respectively. Consequently, the total
force on rhs of (\ref{eqmotion}) reads: 
\begin{eqnarray}
\pmb{F} &=& A \pmb{e}_{z}\times \left (\frac{d\pmb{R}_{V}}{dt} - \pmb{v}_{s} \right ) +
          B \pmb{e}_{z}\times \left (\frac{d\pmb{R}_{V}}{dt} - \pmb{v}_{n} \right ) +
          C \left (\frac{d\pmb{R}_{V}}{dt} - \pmb{v}_{n} \right ) + \nonumber \\
          &+& \pmb{F}_{\textrm{pinning}}\left ( \pmb{R}_{V},\frac{d\pmb{R}_{V}}{dt},\pmb{e}_{z}\times  
          \frac{d\pmb{R}_{V}}{dt} \right ),
\end{eqnarray}
%%%%%%%%%%%%%%%%%%%%%%%%%%%
where the last term, describing the pinning force, breaks Galilean invariance as it corresponds
to the interaction with impurities\footnote{More precisely: the pinning force describes all effects related to the momentum transfer between the vortex and impurities}. It may depend on the location of the vortex, as well as its
velocity \cite{PhysRevB.55.485}. The dependence on the vortex velocity becomes important in the 
limit of large core sizes (as compared to the impurity size), and originates from the
effect of scattering of quasiparticles, bound in the vortex core. It
gives rise to the so-called Kopnin-Kravtsov force \cite{kopninkravtsov, PhysRevB.56.766}.
In the case of the neutron star crust this effect may become important in the region
of pasta phases, where the core size becomes larger (due to weaker pairing
correlations) than the typical scale of spatial density modulation. 
The pinning force is particularly important in the context of neutron star crust \cite{link2009dynamics, haskell2016pinned}, where the pinning-unpinning mechanism is essential for understanding the glitch 
phenomenon~\cite{haskell2015models, haskell2020turbulent}.
The coefficients $A, B, C$ depends on densities $\rho_{s}$ and $\rho_{n}$, describing superfluid and normal components of the system. The equation (\ref{eqmotion})
is obtained by separating terms proportional to $\frac{d\pmb{R}_{V}}{dt}$, and introducing
$\pmb{F}_\textrm{superfluid}$, which depends on $\pmb{v}_{s}$ and $\pmb{v}_{n}$ only (see eg. Ref.~\cite{PhysRevB.57.R8119}).
For the sake of generality we include the vortex mass $M_V$, which, however, is routinely assumed to be
negligible. 
Coefficients $\gamma$ and $\omega$ need to be derived from microscopic theory and are functions of densities $\rho_{s}$ and $\rho_{n}$.
%%%%%%%%%%%%%%%%%%%%%%%%%%%%%%

There are still many uncertainties concerning each term of this equation, particularly for fermionic vortices.
Apart from the Magnus force $\pmb{F}_{M}\propto \rho_{s}\pmb{e}_{z}\times (\frac{d\pmb{R}_{V}}{dt} - \pmb{v}_{s} )$, which has similar form in both bosonic and fermionic case, the form and magnitude of dissipative forces, both transversal and longitudinal, are still debatable~\cite{stone1996,stone2000,sonin2013}.
In the microscopic description, these forces have to be related to quasiparticles and phonon scattering on the vortex core.
It includes the so-called \textit{Iordanskii force}, which is the transverse force
related to the relative motion with respect to the normal component of the superfluid: $\pmb{F}_I \propto \pmb{e}_{z}\times \left (\frac{d\pmb{R}_{V}}{dt} - \pmb{v}_{n} \right )$.
It is predicted to emerge from the scattering of elementary excitations of superfluid on the vortex core~\cite{iordanskii1966}. In the BEC case, it is related to the scattering of phonons, whereas in Fermi systems it would include also quasiparticle
excitations. However the role and the magnitude of this force are still not clear (see eg. the argument in Ref.~\cite{PhysRevLett.79.1321} against the existence
of the transverse force, associated with the normal component). 

Usually, to date, the determination of dissipative forces relied on a semiclassical approach~\cite{kopnin2002, Silaev:2012a}.
It consisted of modeling the presence of the classical Hamilton function: 
$H(L_{z}) = \alpha L_{z}$ grasping the main feature of the chiral branch. 
The evolution of quasiparticles has been modeled through the semiclassical distribution function $f(L_{z}, \phi)$ , where $\phi$ is an angle being canonically conjugate variable to $L_{z}$.
Subsequently, one may search for a solution of Boltzmann eq. describing small deviations from equilibrium distribution. The quasiparticle collisions have been treated in relaxation time approximation, which introduces another parameter to the description.
This approach is, however, expected to work in the deep BCS regime, where one can ignore the presence of discrete states in the band, and the effects related to mini-gap are of secondary importance.

Recently, a microscopic approach rooted in density functional theory has become possible. It provides a  framework for comprehensive studies of static and time-dependent problems including thermal effects. Its popularity in nuclear physics grows with time, and today it is used for describing the properties of nuclei across the whole nuclear chart, nuclear reactions, and properties of neutron star crust~\cite{RevModPhys.88.045004,Col2020,Bulgac2019,magierski2019nuclear}. The modern DFT variants takes into account superfluid correlations, with the pairing field treated as a dynamical variable. One of the variants that proved its usability, especially in the context of ultracold Fermi gases, is known as Superfluid Local Density Approximation (SLDA)~\cite{bulgac2012}. The name refers to the class of the energy density functionals $\mathcal{E}$ that depend only on local densities, such that the energy of system can be written as
\begin{equation}
E =  \int \mathcal{E}[\rho_q(\bm{r}), \nu_q(\bm{r}), \tau_q(\bm{r}), \bm{j}_q(\bm{r}),\ldots]\, d\bm{r}.
\end{equation}
In this expression, $q$ is a generalized index that in the context of nuclear systems refers to protons and neutrons ($q=n,p$), while in the context of cold fermionic gases - to atomic species. The functional is expressed through various densities, where the most common are: normal density ($\rho_q$), kinetic density ($\tau_q$), current density ($\bm{j}_q$) and anomalous density ($\nu_q$). Other densities, not listed here, are indicated by dots. The anomalous density is crucial for the description of superfluidity as it quantifies the presence of Cooper pairs and defines the order parameter $\Delta_{q}(\bm{r})=g\nu_q(\bm{r})$ where $g$ is coupling constant. The order parameter, within this framework, is treated as a complex field, allowing for the description
of quantum vortices. 

The precise form of the functional depends on the considered system; however, for all local functionals that belong to the SLDA class, the energy minimization condition leads to equations that formally have the same structure as Hartree-Fock-Bogoliubov or Bogoliubov-de Gennes equations. For static problems, they have a generic form (spin indices are omitted): 
\begin{align} \label{eq:tdslda2c}
\begin{pmatrix}
h_{q}(\bm{r})  &  \Delta_{q}(\bm{r}) \\
\Delta^{*}_{q}(\bm{r}) & -h^*_{q}(\bm{r}) 
\end{pmatrix}
\begin{pmatrix}
u_{q,n}(\bm{r}) \\
v_{q,n}(\bm{r})
\end{pmatrix}
=
E_{q,n}
\begin{pmatrix}
u_{q,n}(\bm{r})  \\
v_{q,n}(\bm{r})
\end{pmatrix},
\end{align}
where $[u_{q,n},u_{q,n}]^T$ are %two component 
quasiparticle orbitals expressed as mixtures of particles ($v_{q,n}$) and holes  ($u_{q,n}$). 
One has to remember that both $h_{q}(\bm{r})$ and $\Delta_{q}(\bm{r})$ are in fact 
$2\times 2$ matrix in spin space and therefore $[u_{q,n},u_{q,n}]^T$ is in fact
four component vector (see eqs. (\ref{eq:hfbspin})).
They are used to construct densities, e.g., $\rho_q(\bm{r}) = \sum_{E_{q,n}>0} |v_{q,n}(\bm{r})|^2$. 
The single-particle mean-fields $h_q$ and pairing potentials $\Delta_q$, are defined via functional derivatives over densities:
 \begin{align}
 h_q &= - \bm{\nabla}\frac{\delta\mathcal{E}}{\delta\tau_q}\bm{\nabla} 
 + \frac{\delta\mathcal{E}}{\delta\rho_q} 
 - \frac{i}{2} \left\{ \frac{\delta\mathcal{E}}{\delta\bm{j}_q}, \bm{\nabla} \right\}, \\
 \Delta_q &= - \frac{\delta\mathcal{E}}{\delta\nu^*_q},
 \end{align}
where $\left \{\ldots\right \}$ denotes the anticommutator and $\frac{\delta\mathcal{E}}{\delta\bm{j}_q}$
represents a vector constructed by variations over three components of the current $\bm{j}_q$.
One may notice that expressions, like kinetic term $\frac{1}{2m}\nabla^2$, or mean-field potential $U$ are replaced by generalized forms, and by adequately choosing the form of the functional one can go beyond the mean-field approach, while keeping the complexity of the framework at the same level as the mean-field formulation.
This flexibility can be used to create the functional, which
reproduces 
selected observables, derived by other methods or taken from experiments. In the case of ultracold atoms the SLDA functional %assures 
generates the correct equation of state (and related thermodynamic properties), strength of the pairing correlations, and effective mass of quasiparticles~\cite{PhysRevA.76.040502,bulgac2012,PhysRevA.106.013306}. The functionals for nuclear systems are more complex, and they are constructed usually in such a way to reproduce nuclear masses and radii and/or equation of state of neutron matter or symmetric nuclear matter. 
For example, the results presented in Fig.~\ref{fig2} were obtained with Brussels-Montreal Skyrme functional (BSk), which was specifically constructed for astrophysical applications~\cite{chamel2016further,chamel2009further,chamel2008,chamel2009pairing,chamel2010effective}. It assures correct reproduction of the equation of state
and have been fitted to reproduce the magnitude of the pairing gap 
as a function of neutron density, obtained within  Brueckner theory, including
screening effects~\cite{PhysRevC.74.064301}.
Consequently, DFT provides a reliable source of microscopic information about the vortices~\cite{PhysRevLett.96.090403,PhysRevC.75.012805,PhysRevLett.91.190404,PhysRevLett.90.161101,pecak2021properties}.  
The approach can be extended to time-dependent phenomena by replacing $E_{q,k}\rightarrow i\hbar\partial/\partial t$, which allows studying the vortex dynamics~\cite{tylutki2021universal,PhysRevLett.117.232701,barresi2023dissipative,PhysRevA.105.013304,PhysRevA.91.031602,PhysRevLett.112.025301}. Consequently, the time-dependent studies can provide microscopic insight into origin of forces acting on the vortex.

It has been conjectured that an accelerating vortex may be subject to dissipative forces originating from quasiparticles trapped in the core \cite{Silaev:2012a}.
The conjecture, based on a semiclassical approach, presents a mechanism in which the acceleration of a vortex will lead to heating up a gas of quasiparticles in the core. As a result, some of them will be emitted from the core, taking away energy carried by the vortex line.
Although the semiclassical approach is applicable in deep BCS regime, the experiment has been recently performed for a system close to the unitarity, where significant dissipation effects have been observed \cite{Kwon:2021a, Autti2020}. However, the presence of finite temperature did not allow to distinguish the source of dissipation clearly.
Through time-dependent SLDA (TDSLDA) simulations of collisions of two pairs of vortex-antivortex dipoles, it has been possible to identify %the presence of conjectured 
dissipative mechanism due to quasiparticle ejection~\cite{barresi2023dissipative}. The application of the TDSLDA framework revealed that the dissipative mechanism via excitations of the vortex core, while present, emerges to be of secondary importance, and thermal effects dominate the dynamics in experimental realization. 

It has to be noted that, in the context of neutron stars, there is an additional dissipation
channel associated with neutron quasiparticles bound in the vortex core. 
It is due to the interaction
both  with electrons and protons and has been discussed in Ref.~\cite{sedrakian2019}. 
In the case of electrons the coupling is provided by interaction
with the neutron magnetic moment, whereas protons couple strongly through the nuclear interaction.
Detailed dynamics is complicated due to the coupling between electron dynamics and protons~\cite{Haskell2018}.
However, the results indicate that coupling of proton-electron plasma to neutron vortices
lead to short relaxation times implying that magnetars superfluid cores will couple
to the remaining stellar plasma on short dynamical time-scales once proton 
superconductivity is suppressed.
%%%%%%%%%%%%%%%%%%%%%%%%%%%%%%%%%%%%%%%%%%%

 \begin{figure}
 \centering
 \includegraphics[width=0.7\textwidth]{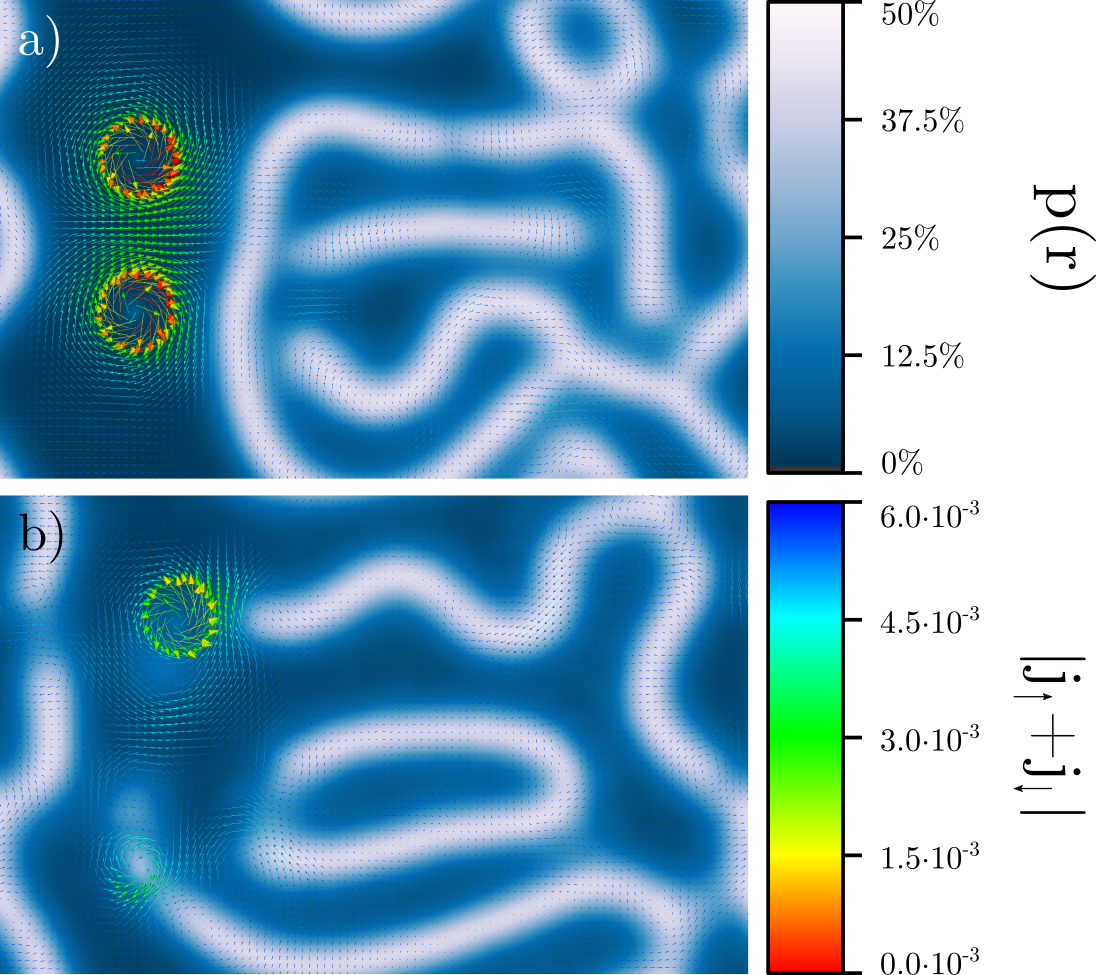}
 \caption{ Vortex dipole (vortex-antivortex pair) 
 propagating in a spin imbalanced unitary Fermi gas (UFG). Spin imbalance induces
 the existence of nodal lines of the pairing field, where the majority spin particles
 are accumulated. Therefore the local polarization is increased along the nodal lines,
 which is seen as white stripes (local polarization is defined as: 
  $p(r)=\frac{\rho_\uparrow-\rho_\downarrow}{\rho_\uparrow+\rho_\downarrow}$). 
 Arrows indicate the local current
 and shows the position of the vortex cores.
 Lower subfigure b) show the configuration after time
 interval $t=250 \epsilon_F^{-1}$ with respect to initial 
 configuration a).}
 \label{fig:polariz}
 \end{figure}

Another successful application of TDSLDA approach has been the determination of the pinning force acting between a vortex and a nucleus immersed in superfluid neutron matter.
To date, studies of this effect were based on comparing the energy of the vortex pinned and unpinned from the impurity~\cite{PhysRevLett.79.3347, PhysRevLett.90.211101, DONATI2004363, DONATI200674, PhysRevC.75.012805, AVOGADRO2008378, seveso2015, klausner2023microscopic}. However, these results were not very accurate due to small energy difference between configurations. One must also remember that the pinning process requires energy dissipation. Otherwise, the vortex will simply orbit 
the nucleus as a result of the Magnus force. Therefore, the proper description of dissipative processes is essential.
Ref.~\cite{PhysRevLett.117.232701} described how the pinning force can be accurately determined by dragging the impurity against the vortex with constant velocity. 
Then, by combining the information about the force with the vortex-nucleus separation, one could extract the force as a function of vortex-nucleus separation.
It was found that the extracted force has a negligible tangential component and is of predominantly central character. 
The effective range of the force is about $10 \textrm{fm}$ for the lower density, increasing to about $15 \textrm{fm}$ for the higher density, consistent with an increasing coherence length with density and decreasing neutron pairing gap. 
The behavior of the total force for small separations demonstrates that it is not merely a function of distance. 
At small separations, the deformation of the vortex line and the nuclear deformation become important degrees of freedom.

\section{Summary and open questions}\label{sec4}

The vortex structure and dynamics discussed in the previous sections are crucial for understanding the role of vortices in neutron stars, particularly in the glitch phenomenon, and more generally, to understand the decay pattern of fermionic quantum turbulence.

Vortex structure is relatively well known, although its modifications due to spin imbalance are still not fully understood.
The situation is different when one considers dynamical properties.
To date, most investigations were performed based on very simplified assumptions, which at most can be valid in deep BCS regimes.
However, the possibility of using the unconstrained time-dependent BdG approach, and more generally, the TDSLDA framework, paved the way to
extract components of forces exerted on a vortex moving through a superfluid.
It whets our appetite concerning the determination of the microscopic underpinning of vortex dynamics in a fermionic environment.
It is particularly important that these studies are performed in close collaboration with experiments in ultracold atoms, which allows the testing of the applicability of the theoretical framework.
To date, the two aforementioned examples of applications of TDSLDA have provided very promising outcomes.

The degree of freedom related to spin imbalance will open a new avenue for studying vortex dynamics. 
The motion of the vortex in the presence of a large number of majority spin particles, may change the vortex dynamics qualitatively. Indeed, their scattering off the core is
expected to enhance dissipative processes considerably~\cite{barresi}.
As an example, in the Fig. \ref{fig:polariz}, the results of TDSLDA simulations
of moving vortex dipole have been shown. In the lower subfigure, one can notice the significant distortion of the initial vortex 
dipole (upper subfigure), which occurs due to interaction
with majority spin particles accumulated in the pairing nodal lines. As a result
some quasiparticles have been trapped in the vortex core.
Therefore, it is expected that
the motion of the vortex in the spin-imbalanced system may shed light on the magnitude of the Iordanskii force, which should be significantly enhanced in this case.
As a result, we may encounter new, unexpected features of vortex dynamics that have not yet been observed.

\bmhead{Acknowledgments}

This  work  was  financially  supported  by  the
(Polish)   National   Science   Center   Grants   under Contracts
No. UMO-2021/43/B/ST2/01191 (PM, AM), UMO-2022/45/B/ST2/00358 (GW, ABa) and 
UMO-2021/40/C/ST2/00072 (DP). This
work used computational resources of Tsubame3.0 supercomputer 
provided by Tokyo Institute of Technology
through the HPCI System Research Project (Project ID: hp230081). 


\end{document}